\documentclass[twocol]{ametsoc}

\journal{jpo}

\bibpunct{(}{)}{;}{a}{}{,}

\title{A model for the wind-driven current in the wavy oceanic surface layer: apparent friction velocity reduction and roughness length enhancement}

\authors{Miguel A. C. Teixeira\correspondingauthor{Department of Meteorology, University of Reading, Earley Gate, PO Box 243, Reading RG6 6BB, United Kingdom.}}

\affiliation{Department of Meteorology, University of Reading}

\email{m.a.teixeira@reading.ac.uk}

\abstract{A simple analytical model is developed for the current induced by 
the wind and modified by surface wind-waves in the oceanic surface layer, 
based on a first-order turbulence closure and
including the effect of a vortex force representing
the Stokes drift of the waves. The shear stress is partitioned between a
component due to shear in the current, which is reduced at low turbulent
Langmuir number ($La_t$), and a wave-induced component, which decays over a
depth proportional to the dominant wavelength. The model reproduces the
apparent reduction of the friction velocity and enhancement of the
roughness length estimated from current profiles, detected in a number of
studies. These effects are predicted to intensify as $La_t$ decreases, 
and are entirely attributed to non-breaking surface waves. The current 
profile becomes flatter for low $La_t$ owing to a smaller 
fraction of the total shear stress being supported by the current shear. 
Comparisons of the model with the comprehensive dataset provided by
the laboratory experiments of Cheung and Street show encouraging 
agreement, with the current speed decreasing as the wind speed increases 
(corresponding to decreasing $La_t$), if the model is adjusted to reflect 
the effects of a full wave spectrum 
on the intensity and depth of penetration of the wave-induced stress. 
A version of the model where the shear stress decreases to zero 
over a depth consistent with the measurements accurately predicts the surface
current speed. These results contribute towards developing physically-based 
momentum flux parameterizations for the wave-affected boundary layer in
ocean circulation models.}

\begin{document}

\maketitle

%

\section{Introduction}
\label{introduction}

Flow coupling across the air-water interface in oceanic regions takes place
within boundary layers where various properties adjust, over a relatively 
small fraction of the depth of the atmosphere and ocean, 
between their values in the 
interior of each fluid. The atmospheric and oceanic surface layers are
the sub-layers of these boundary layers located nearest to the air-water 
interface, which have a decisive importance in mediating the turbulent 
fluxes of various properties (e.g., momentum, heat, gases, pollutants) 
between the atmosphere and the ocean \citep{Csanady_2004}. 

Whereas the atmospheric surface layer over land has a no-slip 
bottom boundary condition applied at the ground, the atmospheric and 
oceanic surface layers in oceanic regions are characterized by continuity of 
velocity and stress at the mobile air-water interface that separates them. 
This, on the one hand, 
leads to the generation of a wind-induced current in the ocean surface layer, 
and on the other hand allows the generation of surface waves at the air-water 
interface. Both of these aspects considerably complicate the physics of these 
surface layers, especially the oceanic one, as is widely recognized 
\citep{Thorpe_2005} and will be further discussed here.

Nevertheless, the oceanic surface layer is still largely understood and modeled 
based on the transposition to the ocean of theories developed for the 
atmospheric surface layer over land, where the effects of surface waves are 
not represented \citep{Kraus_Businger_1994}.   
Deficiencies in this approach become apparent when one realizes that
key parameters in surface layer theory, such as the friction 
velocity $u_*$ and roughness length $z_0$ are deemed to take values in the 
ocean that seem
to be inconsistent with the values of the shear stress and the geometric 
properties of the air-water interface, respectively.

Standard surface layer theory is based on Monin-Obukhov scaling, which in the
limit of neutral stratification reduces to a theory for the 
logarithmic mean wind profile. In the ocean, or in underwater flows measured
in the laboratory, 
such a theory has been applied, with varying degrees of success, to model the
mean current induced by the wind.
However, it has often been detected that the value of $u_*$ 
inferred from the current profile is noticeably smaller than the one that
would be expected from the total shear stress 
(\citealp{McWilliams_etal_1997}, \citealp{Kudryavtsev_etal_2008},
\citealp{Teixeira_Belcher_2010}), a phenomenon sometimes
alternatively interpreted as an increase of the Von K\'arm\'an parameter
(\citealp{Howe_etal_1982}, \citealp{Cheung_Street_1988}, 
\citealp{Craig_Banner_1994}, \citealp{Siddiqui_Loewen_2007}). 
On the other hand, the value of $z_0$
obtained by extrapolating the logarithmic current profile up to the surface
is often much larger than would be expected based on the size
of the surface corrugations deforming the air-water interface, and exceeds
by several orders of magnitude the air-side value of $z_0$
(\citealp{Csanady_1984}, \citealp{Burchard_2001}, \citealp{Soloviev_Lukas_2003},
\citealp{Sullivan_etal_2004}, \citealp{Kudryavtsev_etal_2008}).

There is some awareness that the first aspect is due to the fact that a 
fraction of the surface stress is carried by the surface waves, and therefore
does not support as much shear as if the waves were absent. On the other hand,
the increased values of $z_0$ have been attributed to the effect of surface
waves as roughness elements seen from below, or to wave breaking, but the 
exact mechanism by which this enhancement arises remains rather mysterious. 
The huge disparity between the estimated values
of $z_0$ as seen from the air-side or from the water-side of the air-water
interface is especially puzzling, since the amplitude of the corrugations 
is the same. Even if the flow on both sides of the air-water interface could be
assumed to be aerodynamically smooth, the differences in the value of $u_*$
between air and water would not be enough to explain the magnitude of this
disparity.

\cite{Craig_Banner_1994} and \cite{Craig_1996} developed a model of the
oceanic 
surface layer that produces profiles of the mean current and
of the associated dissipation rate of turbulent kinetic energy (TKE), 
which showed some success in 
predicting both quantities, and was subsequently used and adapted by a number 
of researchers (e.g. \citealp{Drennan_etal_1996}, \citealp{Terray_etal_1999},
\citealp{Gemmrich_Farmer_1999}, \citealp{Burchard_2001}, 
\citealp{Rascle_etal_2006}, \citealp{Feddersen_etal_2007},
\citealp{Rascle_Ardhuin_2009}, \citealp{Gerbi_etal_2009}, 
\citealp{Kukulka_Harcourt_2017}). That model is based on an approximate balance 
between the turbulent fluxes of TKE and 
dissipation, and produces a substantial surface dissipation
enhancement, which is consistent with the
observations of \cite{Gargett_1989}, \cite{Agrawal_etal_1992}, 
\cite{Terray_etal_1996} and \cite{Drennan_etal_1996}. However, it requires
adjusting $z_0$ for each dataset, yielding values of this quantity of
order the height or wavelength of the surface waves, which is much larger
than estimated for an aerodynamically smooth boundary, or from the Charnock
relation. Both \cite{Craig_Banner_1994} themselves and, more recently, 
\cite{Grant_Belcher_2009} recognized that this need to adjust $z_0$ in order
to fit measurements is a weakness of the model.

More recently, \cite{Kudryavtsev_etal_2008} developed a rather elaborate model
based on the momentum and TKE budgets, and assuming a balance between 
turbulence production by
wave breaking and dissipation. This model avoids the strong dependence on
$z_0$ displayed by the model of \cite{Craig_Banner_1994}, but contains many
{\it ad hoc} assumptions and approximations (for example, the parameterization
of the TKE production by wave breaking, or the mixing length definition), 
and nevertheless is so complicated
that the corresponding equations can only be solved numerically. Although
it predicts satisfactorily the qualitative behavior of the mean current 
profiles measured in the laboratory experiments of \cite{Cheung_Street_1988}
and the aforementioned surface dissipation enhancement, it produces 
dissipation profiles that look somewhat artificial and 
seem to underestimate most datasets at small depths (see their Fig. 7).
Although this model succeeds in predicting the 
enhanced values of the apparent $z_0$ in the experiments of 
\cite{Cheung_Street_1988}, it does not explain the reduced values of $u_*$ 
that can also be inferred from the slope of the mean flow profiles.

In this study a very simple model is developed, based on the partition of the 
shear stress in the surface layer between shear-related and wave-related parts,
that reconciles all these results, explaining in particular the discrepancies
between expected and observed values of $u_*$ and $z_0$ in the oceanic surface
layer, purely due to the effect of non-breaking waves (unlike 
\citealp{Kudryavtsev_etal_2008}). 
The model draws heavily on that developed by \cite{Teixeira_2012},
which is inspired by Rapid Distortion Theory (RDT) calculations, and is much 
simpler than the one 
proposed by \cite{Kudryavtsev_etal_2008}, being essentially analytical, 
but produces more accurate results. It has the advantage of being 
formulated as a variant of Monin-Obuhov scaling,
where instead of the Obukhov stability parameter, the key dimensionless
parameters account for the effects of surface waves. These parameters are
the well-known turbulent Langmuir number $La_t$ and (as in Monin-Obukhov theory)
a dimensionless depth, here normalized by the wavenumber of the dominant 
surface waves. An extended version of this model was shown by 
\cite{Teixeira_2012} to give good predictions
of the dissipation rate by comparison with field data from various sources 
(\citealp{Terray_etal_1996}, \citealp{Drennan_etal_1996},
\citealp{Burchard_2001},
\citealp{Feddersen_etal_2007}, \citealp{Jones_Monismith_2008},
\citealp{Gerbi_etal_2009}). The model is tested here by comparison with the
data of \cite{Cheung_Street_1988}, showing good agreement, despite the fact that
(unlike the model of \cite{Kudryavtsev_etal_2008}) it uses a monochromatic 
wave approximation and neglects the viscous boundary layer. 

This paper is organized as follows: section \ref{model} presents the proposed 
model, including its version for a constant shear stress and its extension for
a shear stress that decreases linearly with depth. Section \ref{results} 
contains the results, starting with tests to the model as a function of 
its input 
parameters, and proceeding with its comparison with laboratory data. Finally, 
in section \ref{conclusions}, the main conclusions of this study are summarized.

\section{Theoretical Model}
\label{model}

It will be assumed that the rotation of the Earth and stratification of the
water in the oceanic surface layer can be neglected. The first assumption
is generally acceptable in the surface layer, where the flow is by definition
dominated by turbulent fluxes. The second assumption is acceptable if some
other dynamical process (in the present case the effect of surface waves) is 
stronger than that of buoyancy.
The effect of breaking surface waves will also be neglected. This is a 
working hypothesis, which is not as justifiable as the previous two, but was
shown to be a plausible approximation given the level of agreement achieved 
between the model of \cite{Teixeira_2012} and dissipation data (for further
details concerning its motivation, see that paper). 

The water-side friction velocity $u_*$ and roughness length $z_0$ will be 
specified according to their most fundamental definitions: as the surface 
value of the shear stress,
and as the depth at which the current velocity relative to its surface 
value is zero (without assuming a displacement height), respectively, rather
than based on the slope and intercept of the current profiles (which would be 
misleading in the present context). 

The point of departure for the model is that turbulence in the surface layer
is dominated by the transfer of kinetic energy from the mean wind-driven 
current and the Stokes drift of surface waves to the turbulence, via the
shear production and Stokes drift production terms in the TKE budget 
(see, e.g., \citealp{McWilliams_etal_1997}), which are assumed to be balanced 
locally by the dissipation rate, as in \citeauthor{Teixeira_2011b}
(\citeyear{Teixeira_2011b}, \citeyear{Teixeira_2012}). This balance, although
of questionable accuracy, has been motivated in \cite{Teixeira_2012} by the
TKE budgets presented in the Large Eddy Simulation (LES) studies of
\cite{Polton_Belcher_2007}, \cite{Grant_Belcher_2009} and
\cite{Kukulka_etal_2010} (which did not account for the effects of wave
breaking). More recent supporting evidence for this balance is provided by
\cite{Vanroekel_etal_2012} and \cite{Kukulka_Harcourt_2017}.

RDT studies (e.g. \citealp{Lee_etal_1990}, \citealp{Teixeira_Belcher_2002},
\citealp{Teixeira_Belcher_2010}, \citealp{Teixeira_2011a})
have indicated that the characteristics of the 
turbulence (i.e., its anisotropy and rate of energy transfer from the mean flow)
are determined by its distortion by the mean current shear $dU/dz$ (where
$U(z)$ is the mean current speed), which promotes horizontal `streaky 
structures', and by the Stokes drift gradient $dU_S/dz$
(where $U_S(z)$ is the Stokes drift velocity), which promotes instead 
streamwise vortices with strong vertical velocity fluctuations.
The influence of surface waves can be measured by 
the relative importance of these two strain rates, since the corresponding
production terms in the TKE budget may be written (for a wind stress aligned
in the $x$ direction)
\begin{equation}
-\overline{u'w'} \frac{dU}{dz}, \quad -\overline{u'w'} \frac{dU_S}{dz},
\label{prodtke}
\end{equation}
where $\tau = -\rho \overline{u'w'}$ is the shear stress 
(with $u'$ and $w'$ being the horizontal and vertical turbulent velocity 
fluctuations, respectively) and $\rho$ is the density. It will be assumed
hereafter that $dU/dz$ and $dU_S/dz$ have the same sign ($>0$), which is the
typical situation for wind-driven waves.

\subsection{Scaling of the oceanic surface layer}
\label{scaling}

The vertical gradient of the Stokes drift of a 
deep-water monochromatic surface wave of amplitude $a_w$, wavenumber $k_w$ 
and angular frequency $\sigma_w$ at a depth $z$ 
is given by \citep{Phillips_1977}
\begin{equation}
\frac{dU_S}{dz} = 2 (a_w k_w)^2 \sigma_w {\rm e}^{-2 k_w |z|},
\label{stokes}
\end{equation}
and, to a first approximation, in the surface layer the mean current shear
satisfies
\begin{equation}
\frac{dU}{dz} = \frac{u_*}{\kappa |z|},
\label{current}
\end{equation}
where $\kappa$ is the Von K\'arm\'an constant.
To evaluate the relative importance of the Stokes drift strain rate and mean 
shear
of the current, the ratio of (\ref{stokes}) and (\ref{current}) may be taken
at a representative depth where the flow is affected by surface waves, say 
$|z|=1/(2 k_w)$, yielding
\begin{align}
Ri&=\frac{dU_S/dz}{dU/dz}(|z|=1/(2 k_w))= \kappa {\rm e}^{-1} (a_w k_w )^2 
\frac{c_w}{u_*} \nonumber \\
&= \kappa {\rm e}^{-1} \frac{U_S(z=0)}{u_*} = \kappa {\rm e}^{-1}
La_t^{-2},
\label{ratio}
\end{align}
where $U_S(z=0)=(a_w k_w)^2 c_w$ is the Stokes drift velocity at the surface
and $La_t=(u_*/U_S(z=0))^{1/2}$ is the turbulent Langmuir number.
Incidentally, $|z|=1/(2 k_w)$ is also the depth at which $R$ attains its maximum
(cf. \citealp{Teixeira_Belcher_2010}, \citealp{Teixeira_2011a}).

Consider first the magnitude of $R$ in the atmosphere. Although
one does not often think about Stokes drift in the atmosphere, its magnitude
is similar to that in the ocean, since the wave orbital motions (usually 
submerged in a tangle of turbulent eddies) are likewise of similar magnitude.
$dU_S/dz$ is estimated here as if $dU/dz$ did not affect the
wave motion, which is certainly not strictly true, but provides a 
leading-order approximation. For waves of slope $a_w k_w  \approx 0.1$ and 
wavelengths
in the range $\lambda_w \approx 1-100 \, {\rm m}$, taking into account that 
$k_w=2 \pi/\lambda_w$, then the wavelength is in the range $k_w \approx  
0.06 - 6.3 \, {\rm m}^{-1}$, and using the linear dispersion relation of 
deep-water 
gravity waves, $c_w = \sqrt{g/k_w}$, one obtains $c_w \approx 1.25 -12.5 \, 
{\rm m} \, {\rm s}^{-1}$, with the limits swapped relative to those of 
$k_w$. Taking a typical value of the friction velocity in the atmosphere, 
$u_* \approx 0.3 \, {\rm m} \, {\rm s}^{-1}$,
(\ref{ratio}) yields $R \approx 6 \times 10^{-3}-6 \times 
10^{-2}$ (where $\kappa=0.4$ was assumed), which is very small. 
This means, perhaps unsurprisingly,
that the effect of the Stokes drift in the atmosphere is fairly insignificant,
and the surface layer should be dominated by mean wind shear.

For the oceanic surface layer, although the same estimates for the 
wave characteristics may be used, it must be noted that, to a first 
approximation, the shear
stress $\tau$ is continuous across the air-water interface in steady flow, 
and since by definition $\overline{u'w'}(z=0)=-u_*^2$, then 
$\rho u_*^2$ must be continuous at that interface. Given that the density 
ratio between water and air is $\approx 833$, the friction 
velocity in the water will be smaller by a factor of $\sqrt{833}\approx 29$. 
This gives a typical friction velocity of $u_* \approx 0.01 \, {\rm m} \, 
{\rm s}^{-1}$, yielding
$R \approx 0.17- 1.7$, which is of $O(1)$. In reality, the
value of $u_*$ used in (\ref{current}) should be even smaller, since part of
the shear stress is supported by the wave as well as by the mean shear (as will
be seen later), so that
it is common to have $R$ substantially higher than 1. In addition, it is quite
possible that $a_w k_w > 0.1$, which also increases $R$. This
means that in the ocean it is unacceptable to ignore the 
effect of the Stokes drift of surface waves, and this difference is what 
gives oceanic turbulence its distinctive character, as shown by 
\citeauthor{Teixeira_Belcher_2002} (\citeyear{Teixeira_Belcher_2002},
\citeyear{Teixeira_Belcher_2010}) and \cite{Teixeira_2011a}.

\subsection{Shear stress partition}
\label{partitionsec}

The Craik-Leibovich equations including the effect of the Stokes drift of 
surface waves may be manipulated, in the same way as done for obtaining a TKE
budget including the production terms (\ref{prodtke}), to obtain an equation
for evolution of the shear stress \citep{Teixeira_2011a}
\begin{equation}
\frac{d}{dt} \overline{u'w'} = -\overline{w'^2} \frac{dU}{dz} -\overline{u'^2}
\frac{dU_S}{dz} + {\rm other \,\,\, terms}.
\label{shearst}
\end{equation}
This equation shows that the shear stress receives contributions proportional
to the mean shear and to the Stokes drift strain rate. This prompted 
\cite{Teixeira_2011a} to decompose $\overline{u'w'}$ into shear-induced and
wave-induced components, proportional to the corresponding production terms
explicitly presented in (\ref{shearst}). Hence, the shear-induced component of 
$\overline{u'w'}$ can be parameterized as
\begin{align}
(\overline{u'w'})_s &= \overline{u'w'} \frac{ \overline{w'^2} dU/dz}
{\overline{w'^2} dU/dz + \overline{u'^2} dU_S/dz} = \frac{\overline{u'w'}}
{1 + \frac{\overline{u'^2}}{\overline{w'^2}} \frac{dU_S/dz}{dU/dz}} \nonumber \\
&=\frac{\overline{u'w'}}{1+ \frac{\overline{u'^2}}{\overline{w'^2}}
  2 \kappa (a_w k_w)^2 \frac{c_w}{u_*} k_w |z| {\rm e}^{-2 k_w |z|}},
\label{partition}
\end{align}
where (\ref{stokes}) and (\ref{current}) have been used.
Using the definitions of $u_*$, $U_S(z=0)$ and $La_t$, this may be alternatively
expressed as
\begin{align}
(\overline{u'w'})_s &= -\frac{u_*^2}{1+2 \kappa \frac{\overline{u'^2}}
{\overline{w'^2}} \frac{U_S(z=0)}{u_*} k_w |z| {\rm e}^{-2 k_w |z|}} \nonumber \\
&= -\frac{u_*^2}{1+ 2 \kappa \frac{\overline{u'^2}}{\overline{w'^2}}
La_t^{-2} k_w |z| {\rm e}^{-2 k_w |z|}},
\label{part2}
\end{align}
where it has been noted that in the surface layer the shear stress 
$\overline{u'w'}$ is constant and equal to $-u_*^2$.
If, following \cite{Teixeira_2012}, it is assumed that the quantity
$\gamma= 2 \kappa (k_w |z|) \frac{\overline{u'^2}}{\overline{w'^2}}$ is
approximately constant (which has some plausibility given that $\overline{w'^2}$
must approach zero as $z \rightarrow 0$, particularly in a wave-following
coordinate system -- cf. \cite{Teixeira_Belcher_2002}), then the shear-induced
shear stress takes the form
\begin{equation}
(\overline{u'w'})_s = - \frac{u_*^2}{1+ \gamma La_t^{-2} {\rm e}^{-2 k_w |z|}},
\label{partfin}
\end{equation}
where $\gamma$ is an adjustable (positive) coefficient.
Note that (\ref{partfin}) has the properties of approaching the usual definition
of the total shear stress as either $|z| \rightarrow \infty$ or $La_t
\rightarrow \infty$, both of which make sense physically. The usual wall-layer
scaling for the dissipation rate, consistent with (\ref{current}) and with a
logarithmic current profile, was shown to hold by the observations of various
authors at sufficiently large depths (\citealp{Gargett_1989},
\citealp{Agrawal_etal_1992}, \citealp{Terray_etal_1996}), and is obviously 
recovered when the influence of
surface waves becomes vanishingly small (which corresponds to $La_t \rightarrow
\infty$) (\citealp{McLeish_Putland_1975}, \citealp{Kondo_1976}). 
The remaining part of the shear stress, $\overline{u'w'} - (
\overline{u'w'})_s$, is evidently wave-related, and approaches zero when
either $|z| \rightarrow \infty$ or $La_t \rightarrow \infty$. Its depth of
penetration is clearly, from (\ref{partfin}), of $O(1/(2 k_w))$, although it
should be borne in mind that this particular dependence results directly from
the
monochromatic wave approximation. Other approaches to treat the dependence of
$(dU_S/dz)/(dU/dz)$ (as well as that of $\overline{u'^2}/\overline{w'^2}$) with
depth could result in different functional forms for $(\overline{u'w'})_s$,
with $\gamma$ possibly not being treated as a constant.

An interesting property of (\ref{partfin}) is that, when evaluated at the
surface, it allows the definition of a modified friction velocity affected
by shear, $u_{*s}$, as
\begin{equation}
u_{*s} = -\frac{(\overline{u'w'})_s}{u_*} = \frac{u_*}{1+ \gamma La_t^{-2}}.
\label{modfric}
\end{equation}
Clearly, $u_{*s}$ is always smaller than $u_*$, and can even become much
smaller when $La_t$ is low. This is in agreement with
LES results by, e.g., \cite{McWilliams_etal_1997}, \cite{Li_etal_2005} and
\cite{Grant_Belcher_2009} showing that shear in the current profile
decreases markedly for a constant wind stress $\tau$ as $La_t$ decreases 
(see section \ref{results}).

\subsection{A model for the current profile}
\label{model1}

To obtain a model for the current profile that is consistent with the 
existing surface layer theory, a first-order turbulence closure is applied
to the shear-related shear stress, namely
\begin{equation}
(\overline{u'w'})_s = -K_m \frac{dU}{dz},
\label{1storder}
\end{equation}
where $K_m = \kappa u_* |z|$, as usually defined. Then the shear of the mean
current can be expressed as
\begin{equation}
\frac{dU}{dz} = -\frac{(\overline{u'w'})_s}{\kappa u_* |z|}= \frac{u_*}
{\kappa |z|} \phi_L(La_t,k_w |z|),
\label{sheardef}
\end{equation}
where (\ref{partfin}) has been used in the second equality, and
\begin{equation}
\phi_L (La_t,k_w |z|)= \frac{1}{1+\gamma La_t^{-2} {\rm e}^{-\varepsilon k_w |z|}},
\label{wavephi}
\end{equation}
where $\varepsilon=2$ from (\ref{partfin}), but will hereafter be kept as an 
adjustable parameter for maximum generality.
Note that $\phi_L$ plays in (\ref{sheardef}) a role analogous to that played
by stability functions in Monin-Obukhov theory of the non-neutral surface layer.
The difference resides in the fact that $\phi_L$ depends on wave quantities
(according to (\ref{wavephi})), instead of on stratification.
This formulation is
amenable to improvement, since the form of (\ref{wavephi}) only needs to be
modified to account for missing effects or a more accurate representation of the
effects already considered. The form taken by (\ref{sheardef}) implies that
both at large depths (where usual surface layer scaling is recovered) and near
the surface $z \approx 0$ the current profile is approximately logarithmic,
but with different friction velocities $u_*$ and $u_{*s}$, respectively, 
as expressed by
(\ref{modfric}). The dependence of (\ref{wavephi}) on $z$ is, arguably, the 
simplest possible that benefits from these properties.

To complete the model, it remains to integrate (\ref{sheardef}) between $z=z_0$
(where $U=U_0$, $U_0$ being the Eulerian current at the surface), and a 
generic $z$. This yields
\begin{equation}
U_0-U(z)=\frac{u_*}{\kappa} \int_{z_0}^{|z|} \frac{1}{z'} \frac{1}{1+\gamma 
La_t^{-2} {\rm e}^{-\varepsilon k_w z'}} dz'.
\label{currprof}
\end{equation}
If velocities are normalized by $u_*$ and $|z|$ by $k_w$, (\ref{currprof}) may 
be rewritten
\begin{equation}
\frac{U_0-U(z)}{u_*} = \frac{1}{\kappa} \int_{k_w z_0}^{k_w |z|} \frac{1}{z'}
\frac{1}{1+\gamma La_t^{-2} {\rm e}^{-\varepsilon z'}} dz'.
\label{currprof2}
\end{equation}

Often, current profiles in the surface layer are specified using so-called
wall-coordinates, defined as
$U^+=(U_0-U(z))/u_*$ and $z+= |z| u_*/\nu$, where $\nu$ is the kinematic 
viscosity of water. Using these definitions, (\ref{currprof2}) can be expressed
as
\begin{equation}
U^+ = \frac{1}{\kappa} \int_{\frac{k_w \nu}{u_*} \frac{z_0 u_*}{\nu}}^{\frac{k_w 
\nu}{u_*} z^+} \frac{1}{z'} \frac{1}{1+\gamma La_t^{-2} {\rm e}^{-\varepsilon 
z'}} dz'.
\label{currprof3}
\end{equation}
The advantage of expressing the lower limit of integration in this form is that
for aerodynamically smooth flow, $z_0 u_*/\nu=0.11$ 
(\citealp{Cheung_Street_1988}, \citealp{Kraus_Businger_1994}),
a result that will be used below. The integral in (\ref{currprof2}) or
(\ref{currprof3}) cannot
in general be evaluated analytically. For numerical evaluation purposes only,
it is useful to introduce the further change of variable $z'=\exp{\zeta}$,
which transforms (\ref{currprof3}) into
\begin{equation}
U^+ = \frac{1}{\kappa}
\int_{\log(\frac{k_w \nu}{u_*} \frac{z_0 u_*}{\nu})}^{\log(\frac{k_w \nu}{u_*} z^+)}
\frac{1}{1 + \gamma La_t^{-2} {\rm e}^{-\varepsilon \exp{\zeta}}} d\zeta.
\label{currprof4}
\end{equation}
This eliminates the singularity at $z'=0$, which is especially bothersome for
small values of $z_0$.

In the limit $La_t \rightarrow \infty$, (\ref{currprof4}) 
(or (\ref{currprof3})) can, of course, be
integrated analytically, reducing to
\begin{equation}
U^+ = \frac{1}{\kappa} \log \left( \frac{z^+ \nu}{z_0 u_*} \right) =
\frac{1}{\kappa} \log \left( \frac{|z|}{z_0} \right).
\label{highlat}
\end{equation}
For aerodynamically smooth flow, (\ref{highlat}) further reduces to
\begin{equation}
U^+ = \frac{1}{\kappa} \log \left( \frac{z^+}{0.11} \right) =
\frac{1}{\kappa} \log z^+ + 5.5,
\label{smooth}
\end{equation}
as noted by \cite{Cheung_Street_1988}, where it was assumed that
$\kappa=0.4$.

When plotted with a logarithmic scale for depth, (\ref{currprof4}) consists
of two straight line segments separated by a 
transition depth interval centered around $|z| \approx 1/(\varepsilon k_w)$.
The slope of the current profile in its upper, wave-affected part, is
consistent with the reduced friction velocity $u_{*s}$, given by
(\ref{modfric}), and $u_*$ is of course consistent with the slope of the
profile segment occurring at larger depths (see discussion below). 
The roughness length $z_0$ is
the height at which $U^+=0$, irrespective of whether the current profile is
affected by waves or not. In the latter case, an apparent
roughness length can be defined, which corresponds to the intersect of the
prolongation of the segment of the current profile at large depths with the
axis where $U^+=0$. It can be anticipated that this apparent roughness length
$z_{0w}$ is much larger than the true $z_0$ when the effect of waves is
important, because of the break point (or more precisely transition region)
existing in the current profile.
$z_{0w}$ can be obtained
by integrating (\ref{sheardef}) between $z_0$ and $\infty$ and then
(\ref{current}) back to $z_{0w}$. This yields
\begin{align}
\log(k_w z_{0w}) &= \log(k_w z_0) \nonumber \\
&+ \gamma La_t^{-2} \int_{\log(k_w z_0)}^{\infty}
\frac{{\rm e}^{-\varepsilon \exp{\zeta}}}{1+\gamma La_t^{-2} {\rm e}^{-\varepsilon 
\exp{\zeta}}} d\zeta.
\label{modz0}
\end{align}
Equations (\ref{modfric}), (\ref{currprof4}) and (\ref{modz0}) form the basis 
of the calculations presented in this paper.

It is worth noting that the formulation of the shear stress on which these
equations are based, 
(\ref{1storder}), is strictly local, neglecting any transport effects, 
whereby $dU/dz$ might become negative with $\overline{u'w'}$ remaining also 
negative (corresponding to a negative eddy viscosity in (\ref{1storder})). 
This behavior, which is produced in a number of LES results 
(\citealp{McWilliams_etal_1997}, \citealp{Li_etal_2005}, 
\citealp{Tejada-Martinez_etal_2013}), was recently parameterized by 
\cite{Sinha_etal_2015} by adopting a non-local component of the shear stress, 
akin to those used in 
momentum flux parameterizations for convection. Since the data used in the 
present study \citep{Cheung_Street_1988} do not show such negative current 
shear (a similar example is the top surface layer in Fig. 5
of \cite{Longo_etal_2012}),
that approach is not used here, although it may be viewed as one of the 
possible improvements to the present scheme. 

\subsubsection{Model for a linearly decreasing shear stress}
\label{model2}

For the purpose of comparing the model developed above with the laboratory
measurements of \cite{Cheung_Street_1988} (to be done below), it is convenient
to assume that the shear stress is not constant with depth, but rather varies
linearly from its maximum at the air-water interface to zero at a certain 
depth. This parallels the approach used by \cite{Cheung_Street_1988} to 
estimate the shear stress from their data, and corresponds mathematically
to 
\begin{equation}
\overline{u'w'} = -u_*^2 \left( 1 - \frac{|z|}{\delta} \right) \quad {\rm if}
\quad |z| \le \delta,
\label{linstress}
\end{equation}
where $\delta$ is the depth where $\overline{u'w'}$ becomes zero, and it is 
implied that for $|z|> \delta$, $\overline{u'w'}=0$. In this case, the function
$\phi_L$ must be redefined (for $|z|\le \delta$) as
\begin{equation}
\phi_L \left( La_t, k_w |z|, \frac{|z|}{\delta} \right) 
= \frac{1-\frac{|z|}{\delta}}{1+\gamma
La_t^{-2} {\rm e}^{-\varepsilon k_w |z|}},
\label{redefphi}
\end{equation}
and (\ref{sheardef}) may then be integrated to give
\begin{equation}
U^+ = \frac{1}{\kappa} \int_{\frac{k_w \nu}{u_*} \frac{z_0 u_*}{\nu}}^{\frac{k_w 
\nu}{u_*} z^+} \left( \frac{1}{z'} - \frac{1}{k_w \delta} \right) 
\frac{1}{1+\gamma La_t^{-2} {\rm e}^{-\varepsilon z'}} dz'
\label{currprof5}
\end{equation}
(again valid only for $|z|\le \delta$), replacing (\ref{currprof3}). For
$|z|>\delta$, $U^+=U^+(z^+=\delta u_*/\nu)$, which is a constant.
In the limit $La_t \rightarrow \infty$,
(\ref{currprof5}) reduces to 
\begin{equation}
U^+ = \frac{1}{\kappa} \left[ \log (z^+) - \frac{z^+}{\frac{\delta u_*}{\nu}}
- \log \left( \frac{z_0 u_*}{\nu} \right) + \frac{z_0}{\delta} \right],
\label{logmod}
\end{equation}
which has a log-linear variation and must replace (\ref{smooth}).

Note that, according to (\ref{1storder}) and (\ref{linstress}), 
for $|z|>\delta$, $dU/dz=0$ under
the present assumptions, i.e., 
no mean shear exists and the current speed does not vary. 
This gives the version of the model just described the capability of
predicting the surface value of the wind-induced current (unlike
the version described in the previous subsection, where $U$ varies 
indefinitely). Defining arbitrarily $U(|z|=\delta)=0$, which makes sense 
since this is the value of the current at the depth where the effect of the
surface wind stress is no longer felt, then from the definition of $U^+$ it
follows that $U_0/u_*=U^+(|z|=\delta)=U^+(z^+=\delta u_*/\nu)$, which can be 
obtained from (\ref{currprof5}).

As a caution, it should be emphasized that the assumption of a non-constant 
shear stress, expressed by (\ref{linstress}), is not strictly consistent
with steady and horizontally homogeneous flow (implicit in surface layer 
theory), but hopefully this assumption is still acceptable for 
the present purposes.

\section{Results}
\label{results}

It is instructive first of all to explore the model behavior for a few 
representative cases, because this illustrates in the cleanest possible way 
the range of behavior of the model and its impact on the perceived values
of the water-side values of $u_*$ and $z_0$. More detailed comparisons with 
laboratory experiments follow. In all of these cases, $\gamma$ and 
$\varepsilon$ will be treated as adjustable parameters.

\subsection{Generic behavior of the model}
\label{behaviour}

\begin{figure}[t]
\centering
  \noindent\includegraphics[width=8cm,angle=0]{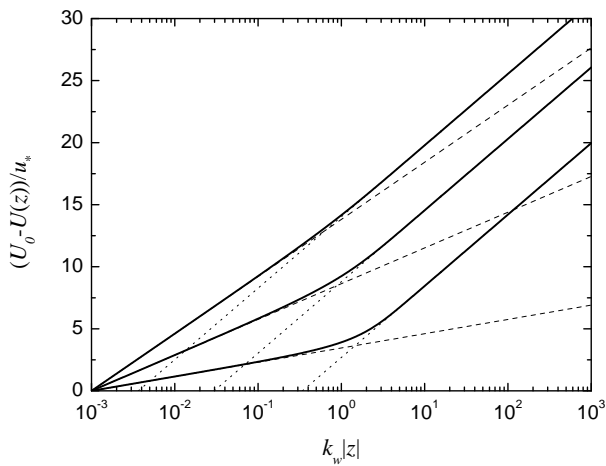}\\
  \caption{Normalized current speed as a function of normalized depth 
for different values of $La_t$, calculated from (\ref{currprof2}) for 
$\gamma=1$, $\varepsilon=1$ and $k_w z_0=0.001$. Solid lines: current profiles, 
for $La_t=2$, $La_t=1$ and $La_t=0.5$ (from top to bottom). Dashed lines:
extension of the asymptotes (with slope $(u_{*s}/u_*)/\kappa$) corresponding 
to the currents at small depths to large depths. Dotted lines: extension of 
the asymptotes (with slope $1/\kappa$) corresponding to the currents at 
large depths up to the depths where the currents would be zero, corresponding 
to the values of the apparent roughness length $k_w z_{0w}$.}
\label{f1}
\end{figure}

Figure \ref{f1} shows profiles of $U^+$ as a function of $k_w |z|$ from 
(\ref{currprof2}) for
$k_w z_0=0.001$ and different values of the turbulent Langmuir number
$La_t=0.5, 1, 2$, assuming that $\gamma=1$ and $\varepsilon=1$, for simplicity.
Note that these values of $\gamma$ and $\varepsilon$ are of the same order
of magnitude as those adopted by \cite{Teixeira_2012}. The results are not
qualitatively very sensitive to $k_w z_0$, and variation of this parameter
merely leads to a rescaling of the horizontal axis, with a narrower transition
region between the two logarithmic portions of the curves occurring 
for values of $k_w z_0 \ll 1$ . 

$La_t=2$ intends to represent shear-dominated turbulence, $La_t=0.5$ 
Langmuir (i.e., wave-dominated) turbulence, and $La_t=1$ turbulence with a 
transitional character.
As can be seen in Fig. \ref{f1}, the current profiles (denoted by the solid
curves) have a lower portion
with invariant slope for larger depths. 
This slope, when expressed in terms of $U^+/\log |z|$,
is $1/\kappa$, because of the way $U^+$ is normalized. At smaller depths
the current profile has a lower slope (prolonged to larger $|z|$ as the 
dashed asymptotes), 
which is proportional to the values of the ratio $u_{*s}/u_*$ in each case. 
From (\ref{modfric}) (for $\gamma=1$), these values 
are $u_{*s}/u_*=0.8$ for $La_t=2$, $u_{*s}/u_*=0.5$ for $La_t=1$ and
$u_{*s}/u_*=0.2$ for $La_t=0.5$. On the the other hand, if the lower portion
of the current profile is prolonged towards the surface (dotted line 
asymptotes), one obtains an ``effective'' value of the roughness length,
expressed by (\ref{modz0}), which would be obtained by
ignoring the upper portion of the current profile. 
For $La_t=2$, $k_w z_{0w}=0.004$, for $La_t=1$,
$k_w z_{0w}=0.030$ and for $La_t=0.5$, $k_w z_{0w}=0.341$, which shows 
dramatically how $z_{0w}$ may become various orders of magnitude larger 
than $z_0$ as $La_t$ decreases (see further discussion below).

Note that, according to the present model, 
if measurements are taken at a range of depths well below the transition region
located around $|z| \approx 1/(\varepsilon k_w)$,
the friction velocity corresponding to the total momentum flux $u_*$ will
be diagnosed correctly from the current profile, but the roughness length 
$z_0$ will be strongly overestimated as $z_{0w}$. Conversely, if measurements 
are taken at a range of depths above this transition region 
(if that is feasible), 
$z_0$ will be correctly diagnosed from the current profile, but
$u_*$ will be underestimated as $u_{*s}$. Data taken from an intermediate 
depth range coinciding with the transition between the two asymptotic 
portions of the profile (if they form a reasonably straight line 
in a logarithmic scale) will lead both to an overestimation of $z_0$ and to an 
underestimation of $u_*$. It is likely that at least one of these three
possibilities occurs in a large fraction of the available field or laboratory 
measurements of wave-affected mean currents.

Circumstantial evidence that this is so is provided by the reported
need to change (more specifically decrease) the value of the Von K\'arm\'an 
constant to achieve
an adequate collapse of measured current profiles in wall coordinates 
(\citealp{Howe_etal_1982}, \citealp{Cheung_Street_1988}, 
\citealp{Craig_Banner_1994}, \citealp{Siddiqui_Loewen_2007}), unless the
friction velocity used to define $U^+$ is that diagnosed from the current
profile itself, here defined as $u_{*s}$ \citep{Siddiqui_Loewen_2007}, which
masks this problem. 
Clearly, neither of these procedures is very satisfactory, given its 
arbitrariness. More evidence supporting the discussion in the preceding 
paragraph is provided by the consistently high reported
values of the roughness length diagnosed from current profiles, exceeding
by orders of magnitude the value that would be expected from the morphology
of the air-water interface, or the flow regime (\citealp{Csanady_1984}, 
\citealp{Burchard_2001}, \citealp{Soloviev_Lukas_2003},
\citealp{Sullivan_etal_2004}, \citealp{Kudryavtsev_etal_2008})). Yet more 
indications, of a more doubtful but suggestive nature, are provided by
the fact that the slope of wave-affected currents plotted in wall-layer
coordinates increases in some cases at larger depths (see, e.g., the
diamond and circle symbols in Fig. 1 of \cite{Cheung_Street_1988}, or 
the black circles and diamonds in Fig. 6 of \cite{Siddiqui_Loewen_2007}).

Although both a decrease of the friction velocity and an increase of the
roughness length, as diagnosed from current profiles, might be expected
as a result of vertical mixing of momentum due to wave breaking, 
the remarkable property of the model proposed here is that this phenomenon 
arises
simply due to the partition of the shear stress imposed by non-breaking waves,
something that can be traced back to the production terms of the shear
stress budget (\ref{shearst}), and is thus much easier to pinpoint physically. 
It is, of course, possible, and even likely, that both processes act in 
concert when wave breaking does occur, but it is striking that the present 
mechanism does not require wave breaking.

\begin{figure}[t]
\centering
  \noindent\includegraphics[width=7.5cm,angle=0]{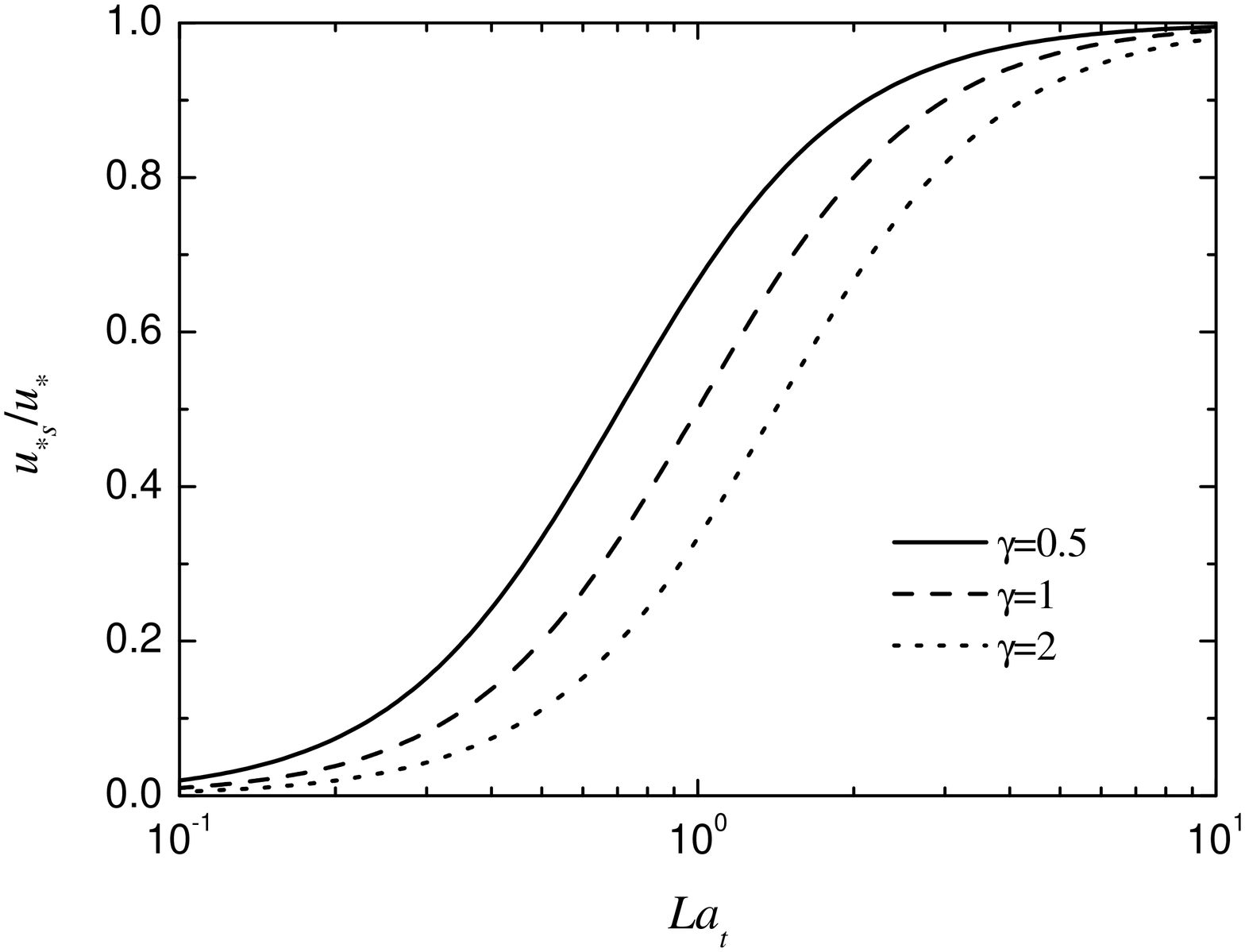}\\
  \caption{Ratio of the shear-associated friction velocity to the total 
friction velocity as a function of $La_t$ for different values of $\gamma$, 
calculated from (\ref{modfric}). See legend for meaning of different line 
types.}
\label{f2}
\end{figure}
Figure \ref{f2} shows the variation of $u_{*s}/u_*$ as a function of $La_t$ 
for different values of the calibrating constant $\gamma$, from (\ref{modfric}).
Unsurprisingly, this ratio takes values that range from $\approx 1$ for large
$La_t$ to $\ll 1$ for small $La_t$. Clearly, what matters for a correct
representation of the variation in between is the value of $\gamma$, with
large values corresponding to strong wave effects and
small values to weaker wave effects. This partition of
the friction velocity, or between the corresponding shear-induced and 
wave-induced stresses, is not an often measured or calculated quantity, but 
Fig. 5 of \cite{Bourassa_2000} presents an example with some relevance, even
if a quantitative comparison is not easy. If an increase in wind 
speed is equated with a decrease of $La_t$ (an idea that is confirmed by 
the comparisons of the next subsection), and the ratio of 
the aqueous shear
stress to the total atmospheric stress is equated with $u_{*s}/u_*$
(which must at least be partially correct because the aqueous stress is 
estimated
from current profiles), the decreasing trend of this ratio with increasing
wind speed in Fig. 5 of \cite{Bourassa_2000} is consistent with Fig. \ref{f2}.  

\begin{figure}[t]
\centering
  \noindent\includegraphics[width=7.5cm,angle=0]{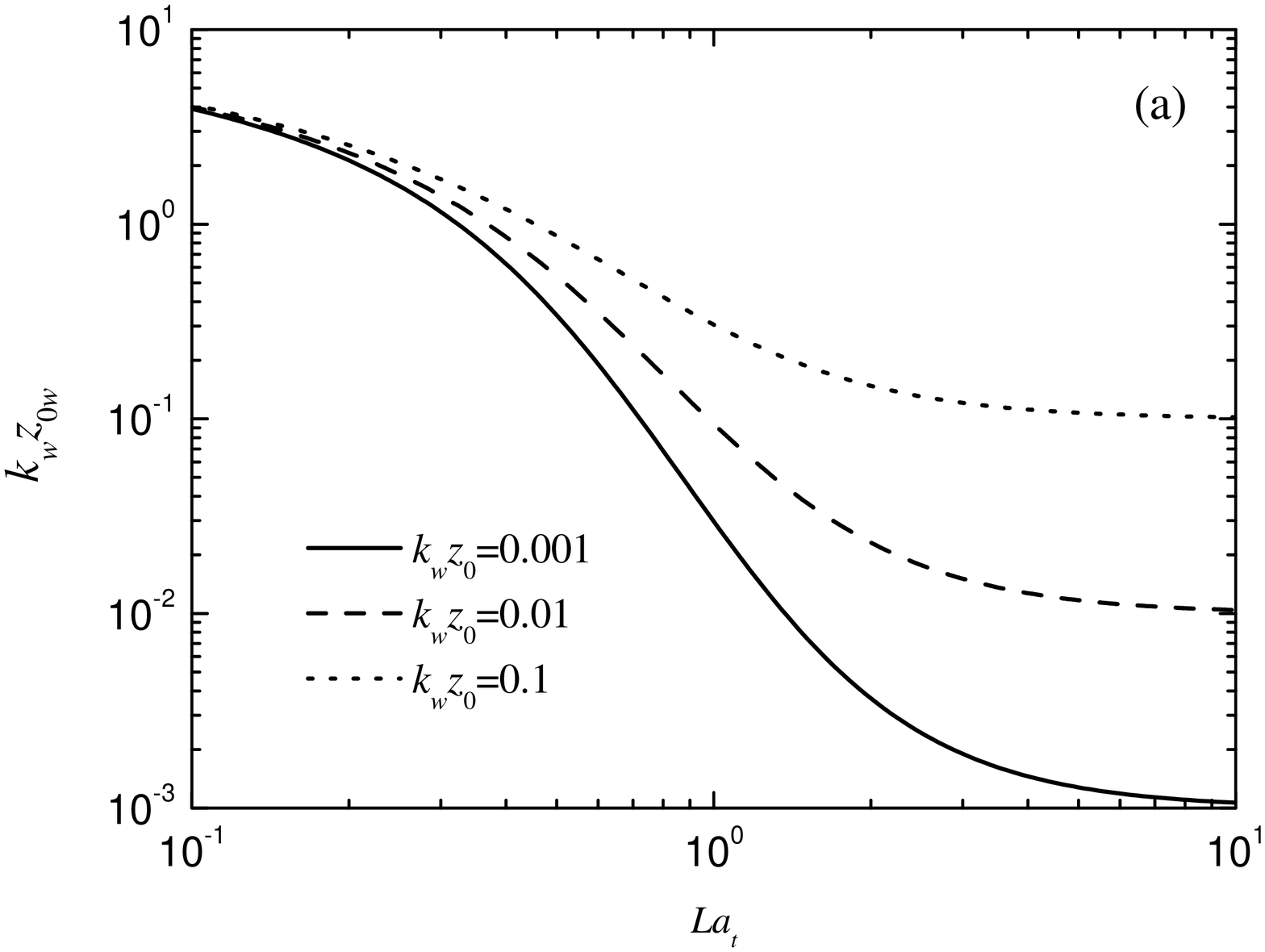}\\
  \noindent\includegraphics[width=7.5cm,angle=0]{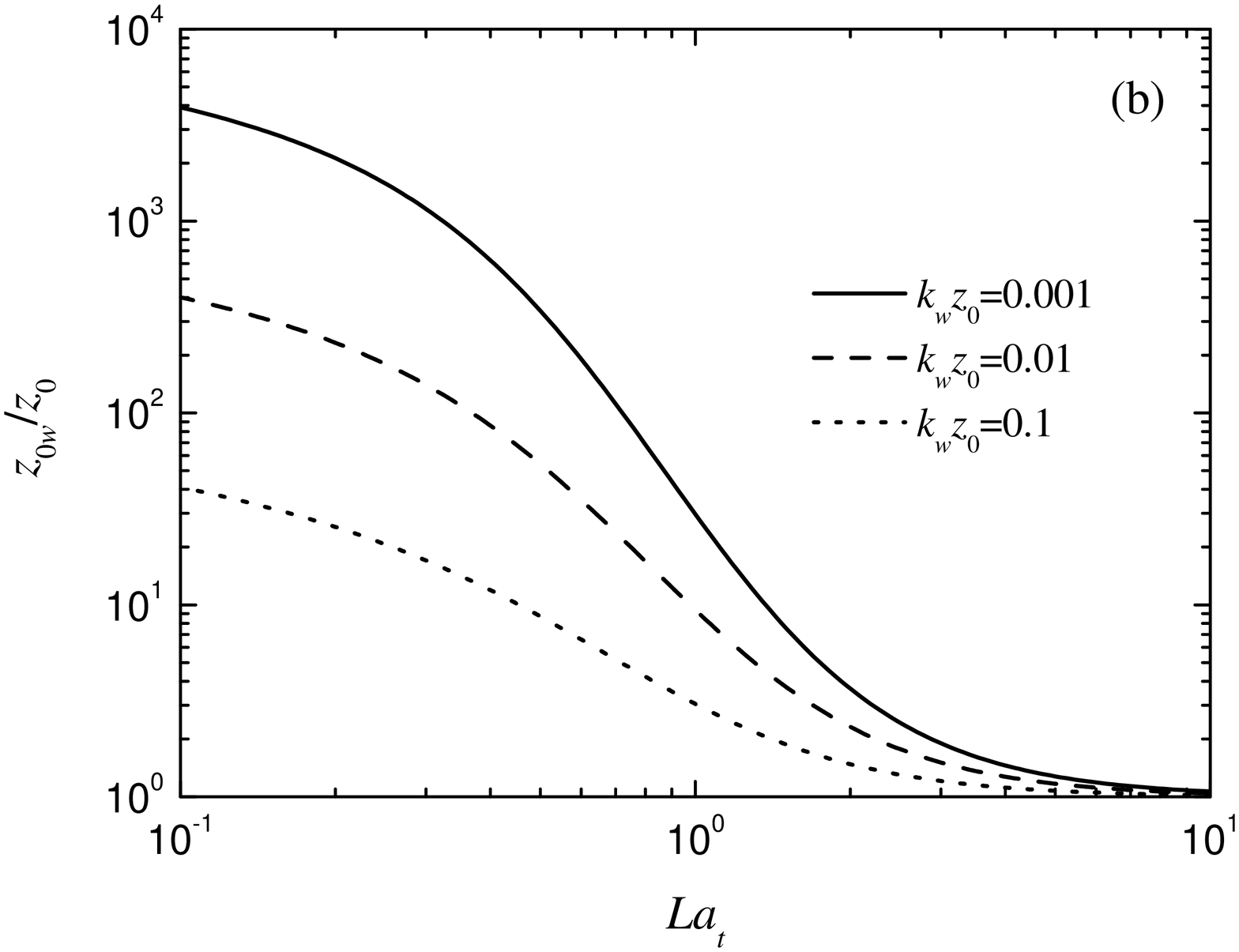}\\
  \caption{Normalized apparent roughness length 
as a function of $La_t$ for $\gamma=1$ and $\varepsilon=1$ from (\ref{modz0}), 
for different 
values of $k_w z_0$. (a) Apparent roughness length normalized by $k_w$, 
(b) ratio of apparent to true roughness length. 
See legend for meaning of different line types.}
\label{f3}
\end{figure}
Figure \ref{f3} presents the variation of $k_w z_{0w}$ and $z_{0w}/z_0$ 
as a function of $La_t$
from (\ref{modz0}) for $\gamma=1$ and $\varepsilon=1$ (as assumed in Fig. 
\ref{f1}) and 
different values of $k_w z_0$. As expected, $k_w z_{0w}$ approaches $k_w z_0$
for large values of $La_t$, but tends to a value independent of $k_w z_0$
at small $La_t$. What this means is that at low $La_t$ $z_{0w}$ scales with 
$k_w^{-1}$
rather than with $z_0$. This is confirmed by the ratio $z_{0w}/z_0$, which 
only approaches 1 for large values of $La_t$, whereas
it tends to be very high for small $La_t$. As is consistent with the behavior
of $k_w z_{0w}$, $z_{0w}/z_0$ at low $La_t$ is inversely
proportional to $k_w z_0$. Since in real situations $k_w z_0$ may easily 
be as small as $10^{-5}$, the amplification of the apparent roughness
length can be very pronounced. A qualitative comparison with Fig. 3
of \cite{Bourassa_2000} is pertinent. Although the dependence of $z_0$
(which should be taken as $z_{0w}$ in the present notation) with $u_*$
in that figure cannot be tested quantitatively because wave information 
is missing, and the
dependence on $u_*$ affects both the true value of $z_0$ (see (\ref{charnock})
below) and (\ref{modz0}) via the definition of $La_t$, the important point
to retain from Fig. 3 of \cite{Bourassa_2000} is the enormous amplification of
$z_0$. \cite{Bourassa_2000} notes that $z_0$ is about $10^5$ larger than
expected from Charnock's relation (and therefore much higher than the values
estimated for the true $z_0$ in the next subsection).

\subsection{Comparison with Cheung {\it et al.} (1988)}
\label{comparison}

Finding adequate datasets to test the present model is challenging,
because usually the quantities required as input to the model are not 
measured. First of all, measuring current profiles in the field with the
required accuracy is extremely difficult, hence the most relevant studies 
typically involve laboratory experiments. Even in those cases, almost invariably
not all relevant wave quantities are measured (\citealp{Bourassa_2000},
\citealp{Siddiqui_Loewen_2007}, \citealp{Longo_etal_2012}),
and often the shear stress is not measured directly,
but rather estimated from the current profiles (\citealp{Bourassa_2000},
\citealp{Siddiqui_Loewen_2007}), which makes comparisons more difficult 
(the erratic behavior of the current speeds measured
by \cite{Siddiqui_Loewen_2007} as a function of the wind speed is another
reason to exclude their data).
A notable exception are the laboratory experiments of
\cite{Cheung_Street_1988} of the current beneath surface waves generated by the
wind. The relevant quantities are presented in their Table 1. 
As \cite{Kudryavtsev_etal_2008} do for the comparison presented in their Fig. 
10, only wind-generated waves are considered here and the case among these
waves with the lowest wind-speed (where the wave amplitude is so small as
to be barely measurable) is ignored.

The experiments with mechanical waves are excluded from this comparison because
the assumption of the model that 
$dU/dz$ and $dU_S/dz$ have the same sign may not be 
strictly satisfied. The possibility that $dU/dz$ and $dU_S/dz$ have 
opposite signs has been demonstrated 
by \cite{Pearson_2018}, for situations with weak (or no) wind, when turbulence 
exists beneath a wave field. This leads to a suppression
of the instability to Langmuir circulations (which requires 
$(dU/dz)(dU_S/dz)>0$), modifying
the stress partition assumed in (\ref{partfin}), which relies on the 
existence of that instability \citep{Teixeira_2011a}. 

For a reasonable range of input parameters, the present model predicted 
almost no difference between the current profiles beneath wind waves
for the two lowest wind speeds in Table 1 of \cite{Cheung_Street_1988}. 
This justifies (following \cite{Kudryavtsev_etal_2008}) 
ignoring the profile for the lowest wind speed, $1.5 \,{\rm m}
\, {\rm s}^{-1}$, which has a roughness 
length smaller than that expected for an aerodynamically smooth flow, and 
might be affected by some inaccuracy.

\begin{figure}[t]
\centering
  \noindent\includegraphics[width=8cm,angle=0]{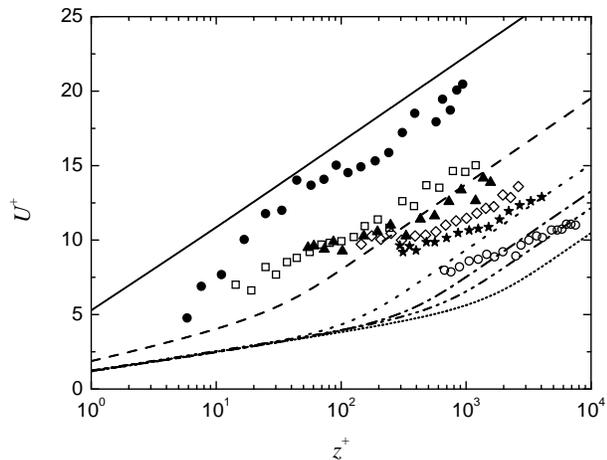}\\
  \caption{Comparison between normalized current speed profiles 
in wall-coordinates from the model developed here, given by 
(\ref{currprof3}) or (\ref{currprof4}) (lines), and from the 
measurements of \cite{Cheung_Street_1988} (symbols), for different wind speeds. 
The model assumes $\gamma=2$, $\varepsilon=2$, and $c_1=0.11$, $c_2=0$ 
in (\ref{charnock}). Solid line and filled circles: 
$2.6 \,{\rm m}\,{\rm s}^{-1}$, dashed line and squares: 
$3.2 \,{\rm m}\,{\rm s}^{-1}$, dotted line and triangles: 
$4.7 \,{\rm m}\,{\rm s}^{-1}$, dash-dotted line and diamonds:
$6.7 \,{\rm m}\,{\rm s}^{-1}$, dash-double-dotted line and stars:
$9.9 \,{\rm m}\,{\rm s}^{-1}$, short-dotted line and open circles:
$13.1 \,{\rm m}\,{\rm s}^{-1}$.}
\label{f4}
\end{figure}

The first comparison to be made uses an uncalibrated version of the 
infinite-depth model described in section \ref{model}\ref{model1}.
The values of $u_*$ from Table 1 of \cite{Cheung_Street_1988} are used 
directly in the model, the wave orbital velocity $a_w k_w c_w$ is taken as 
$(\overline{\tilde{u}_0^2})^{1/2}$, its angular frequency
$\sigma_w$ is equaled to $2 \pi f_D$, and the corresponding wavenumber is 
$k_w=\sigma_w^2/g$ from the linear
dispersion relation of deep-water surface waves. 
An evidently crucial detail is how to 
define $z_0$. As a first approximation the definition valid for aerodynamically
smooth flow is adopted: $z_0=0.11 \nu/u_*$ \citep{Kraus_Businger_1994},
with $\nu= 10^{-6}\,{\rm m}^2 \,{\rm s}^{-1}$.
Figure \ref{f4} shows a comparison of the model with the data 
presented in Fig. 1 of \cite{Cheung_Street_1988} (excluding the upward pointing
triangles for the reasons explained above), assuming $\varepsilon=2$ and 
$\gamma=2$, as in \cite{Teixeira_2012}.

It can be seen in Fig. \ref{f4} that the behavior of the measured currents
is fairly well reproduced qualitatively, with a decrease of the overall 
current speed as the wind speed increases. In terms of the input parameters
of the model, this is due to a decrease of the turbulent Langmuir number
$La_t$ as the wind speed increases. Noteworthy disagreements are that the range
of variation of the current speed in the model is too wide compared with the 
data, in particular, the current speed in wall coordinates is overestimated
for the lowest wind speed and underestimated for the highest ones. 
Additionally,
although two logarithmic portions of the current profile exist in the model at 
the
highest wind speeds (lowest values of $La_t$), these portions to not coincide
with the data that show a reduced slope (e.g., stars and open circles).
Finally, the detailed variation with the wind speed is not reproduced. 
While most of the variation occurs at the lowest wind speeds in the 
model and weakens roughly monotonically as $La_t$ decreases, the rate of 
variation seems to increase again at the highest wind speeds in the data.

When Fig. \ref{f4} is compared with Fig. 10 of \cite{Kudryavtsev_etal_2008}, 
it may be
noticed that the agreement is only marginally less satisfactory, 
with the exception of 
the current profile for the lowest wind speed, where consideration of the 
effect of the viscous boundary layer by \cite{Kudryavtsev_etal_2008} 
substantially improves 
the agreement at small depths. Curiously, this model 
has rather similar deficiencies as the present one, 
namely an overestimation of the sensitivity to the wind speed 
at intermediate values of that parameter and, on the contrary, a too weak 
dependence for its highest values. 
The model of \cite{Kudryavtsev_etal_2008} is naturally unable 
to capture
the apparent reduction of $u_*$ by the wave stress, but a somewhat similar 
effect is mimicked by the transition of the profiles to their viscous 
boundary layer form.

Clearly, the comparison presented in Fig. \ref{f4} indicates an 
overestimation of parameter $\gamma$ in the present model. One might wonder
why this happens, given that this calibration seemed to work for predictions
of the dissipation rate by \cite{Teixeira_2012}, and also in his preliminary
calibration procedure using current profiles from the LES of 
\cite{Li_etal_2005}. Possible reasons are speculative, but might have to 
do with inadvertently accounting for the effect of wave breaking 
in the first case, and adopting a value of $\gamma$ suitable for monochromatic
waves in the second, both conditions which are not applicable here. It seems 
fortuitous that both of these distinct differences should lead to a similar 
value of $\gamma$.

In order to improve agreement with the data of \cite{Cheung_Street_1988}, 
$\gamma$ and $\varepsilon$ may be readjusted, but this turns out not to 
be sufficient. It is likely that
the flow in the experiments under consideration was not always aerodynamically
smooth, but rather aerodynamically rough at the highest wind speeds, because
of the small-scale corrugations forced at the air-water interface by the wind
stress. A form of the roughness length that reflects this is
\begin{equation}
z_0 = c_1 \frac{\nu}{u_*} + c_2 \frac{u_*^2}{g},
\label{charnock}
\end{equation}
where $c_1$ and $c_2$ are coefficients, and the second term is of a form 
analogous to the Charnock relation, but using the friction velocity 
in the water. Here, $\gamma$, $\varepsilon$, $c_1$ and $c_2$ are adjusted
to produce the best possible agreement with the data of 
\cite{Cheung_Street_1988}. The values found for the infinite-depth model are
$\gamma=0.5$, $\varepsilon=0.5$, $c_1=0.2$ and $c_2=0.9$.

\begin{figure}[t]
\centering
  \noindent\includegraphics[width=8cm,angle=0]{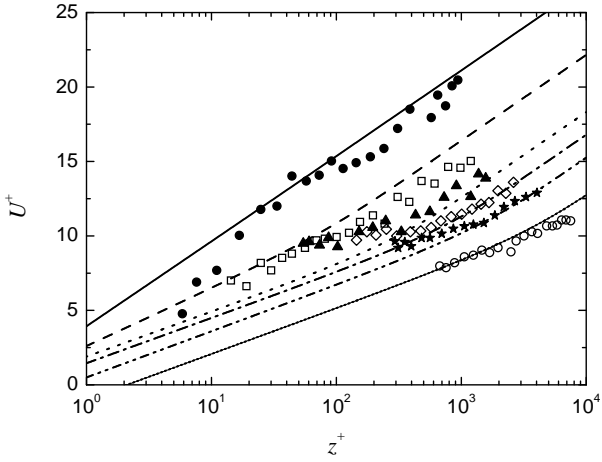}\\
  \caption{Similar to Fig. \ref{f4}, but for $\gamma=0.5$, $\varepsilon=0.5$,
$c_1=0.2$ and $c_2=0.9$.}
\label{f5}
\end{figure}

Figure \ref{f5} shows a comparison of the model with the data of
\cite{Cheung_Street_1988} using
these adjusted parameters. The agreement is much better than in Fig. \ref{f4},
in particular for the extreme wind speed values considered (this is not 
surprising, being a result of the calibration procedure). Agreement is 
less close for the lowest wind speed considered at small depths, 
due to the absence of a viscous boundary layer in the model, but this is a 
minor limitation. The transition of
the datasets from a slope corresponding to $u_*$ to the smaller value 
corresponding to $u_{*s}$ is fairly well reproduced, 
occurring somewhere around between the wind speeds of 4.7 and $6.7\,{\rm m}
\, {\rm s}^{-1}$. However, at these intermediate wind speeds, the current at
the smallest depths covered by the data 
is somewhat underestimated by the model (the shear suggested by 
the data at those depths is weaker than expected). Additionally while 
the current is slightly underestimated for a wind speed of $4.7 \,{\rm m}\,
{\rm s}^{-1}$, it is on the contrary slightly overestimated for a wind speed 
of $3.2 \,{\rm m}\,{\rm s}^{-1}$. It is perhaps risky to attach too much 
relevance to these discrepancies, given the limited precision of the 
measurements (which are, nevertheless, among the most precise that could be
found). 

The value of $\gamma$ was already discussed above. The value of $\varepsilon$
adopted for this comparison would correspond to the Stokes drift of a 
monochromatic wave with a wavelength 4 times larger than the wavelength of the 
dominant waves, obtained from the data.
The significance of this mismatch for non-monochromatic waves (such as the
ones under consideration) is not obvious, but indicates a larger depth of
penetration of the wave-induced stress than would be expected.
The Stokes drift gradient of a wave spectrum is known to be characterized by
a larger penetration depth than a monochromatic wave with the same dominant
wavelength (Fig. 18 of 
\cite{Li_Garrett_1993}, and this may perhaps account for a similar 
effect on the wave-induced stress.

Concerning parameters estimated for (\ref{charnock}), $c_1=0.2$ is substantially
larger than the value of 0.11 most commonly accepted for aerodynamically 
smooth flow. It is worth noting that, in Fig. 10 of \cite{Kudryavtsev_etal_2008}
the thin line (corresponding to aerodynamically smooth flow) assumes
$z_0=0.18 \nu/u_*$, which is not too different from the value employed here.
Regarding $c_2$, the Charnock relation, when expressed in terms of
the friction velocity in the airflow, usually has a coefficient of 0.015.
Taking into account continuity of the shear stress
at the air-water interface, when that relation is expressed in terms 
of the 
friction velocity in the water the coefficient should become $833 \times 0.015
=12.5$. This is clearly much larger than $c_2=0.9$ used here, 
but it should be noted
that the Charnock relation, as usually formulated, is valid in the open 
ocean and for a fully developed wave field, which are very distinct conditions 
from those produced in the experiments of \cite{Cheung_Street_1988}. 
Additionally,
continuity of the shear stress at the air-water interface (used in the above
calculation) assumes 
equilibrium, which is not warranted in these experiments either. 
Nevertheless, a reassuring aspect
is that, on dimensional grounds, the quantities included in 
(\ref{charnock}) are still likely to be the most relevant.   

\begin{figure}[t]
\centering
  \noindent\includegraphics[width=8cm,angle=0]{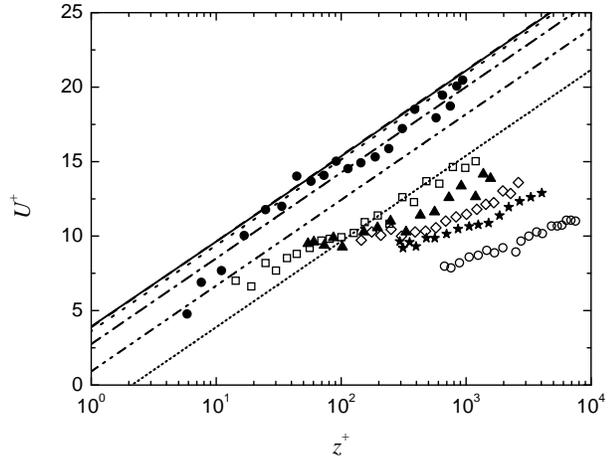}\\
  \caption{Similar to Fig. \ref{f5}, but for $\gamma=0$ (i.e., no wave 
effects).}
\label{f6}
\end{figure}
It might be argued that the agreement between model and measurements in 
Fig. \ref{f5} was artificially improved by allowing $z_0$ to vary 
according to (\ref{charnock}). To test this, 
Fig. \ref{f6} shows a similar comparison,
but where wave effects are ignored altogether, and only the dependence of 
$z_0$ on $u_*$ via (\ref{charnock}) is retained (with similar values of $c_1$ 
and $c_2$). It is clear that this
dependence, by itself, is unable to produce a satisfactory agreement
with the 
measurements, particularly at the highest wind speeds, and naturally does
not represent the decrease of the apparent value of $u_*$, although it does
represent a part of the increase of $z_0$ required to match the data.

\begin{figure}[t]
\centering
  \noindent\includegraphics[width=8cm,angle=0]{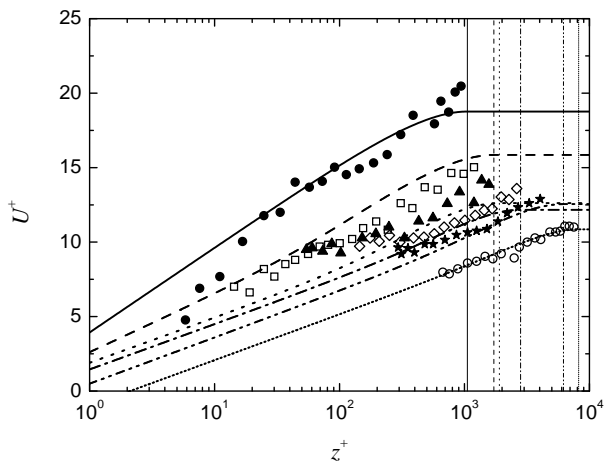}\\
  \caption{Similar to Fig. \ref{f5}, but using the limited-depth model for which
 the shear stress decreases linearly with depth (\ref{currprof5}), with
$\gamma=0.5$, $\varepsilon=1$, $c_1=0.2$ and $c_2=0.9$. The vertical lines
(same type as the corresponding current profiles) denote the depths at which 
the shear stress reaches zero in each case ($z^+=\delta u_*/\nu$).  Note that 
the current profiles for larger depths remain constant.}
\label{f7}
\end{figure}

Figure \ref{f7} shows a similar comparison to Fig. \ref{f5}, but using the
finite-depth model developed in section \ref{model2}. 
Because of the log-linear form of the current profile, 
the current
solutions are no longer composed of straight line segments when
using a logarithmic depth 
scale, but tend to have a reduction in shear at the depths near
where the shear stress becomes zero (and the current speed stabilizes), 
marked by the vertical lines in Fig. \ref{f7}. 
The parameter values used in Fig.
\ref{f7} were $\gamma=0.5$, $\varepsilon=1$, $c_1=0.2$ and $c_2=0.9$.
The agreement between the model and measurements is roughly
as satisfactory as in Fig. \ref{f5}, with essentially the same deficiencies
in the mid-range of wind speeds. At the largest depths considered (near to 
$|z|=\delta$) the model tends to underestimate the measurements more, perhaps 
because the reduction of shear in those regions is overestimated by the 
assumption of a linearly decreasing shear stress. In reality, the fact that the 
shear stress decays to zero more gradually might explain why no marked 
reduction in the shear is detectable in the data at those depths. The
existence of this shear reduction in the model counteracts the 
transition to a larger shear that occurs below the depth 
$|z|\approx 1/(\varepsilon
k_w)$, when this fortuitously coincides with $|z|=\delta$. This is what allows 
a larger value of $\varepsilon$ to be employed in Fig. \ref{f7}.

\begin{figure}[t]
\centering
  \noindent\includegraphics[width=7.5cm,angle=0]{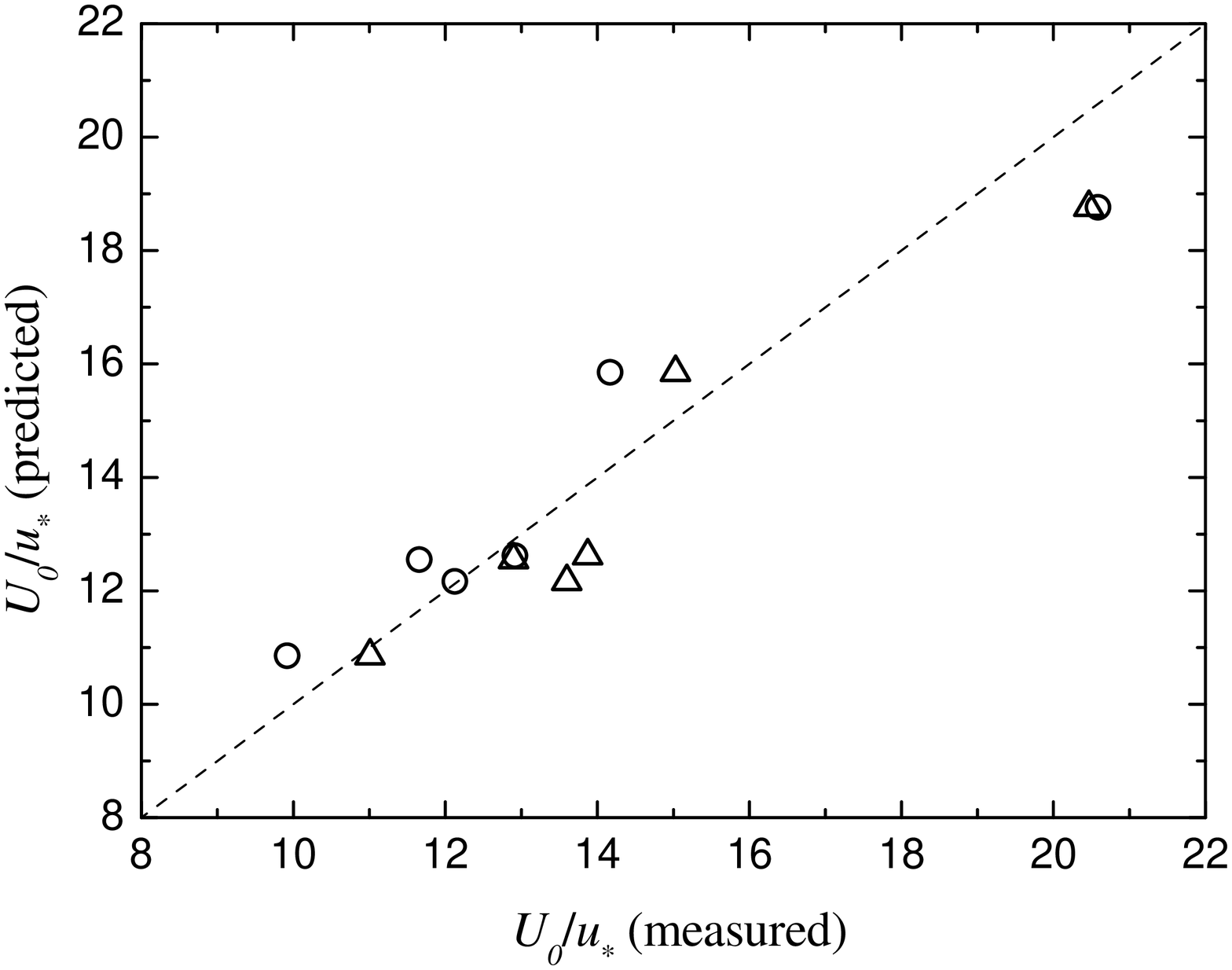}\\
  \caption{Normalized current speed at the surface predicted by the 
limited-depth model (\ref{currprof5}) (for $z+=\delta u_*/\nu$) as a function
of corresponding values derived from the measurements. Circles: measured values
taken directly from Table 1 of \cite{Cheung_Street_1988}, triangles: measured
values taken as the data point at the largest depth from the datasets 
corresponding to each different wind speed.} 
\label{f8}
\end{figure}

A noteworthy property of this finite-depth model is that it enables an
estimation of the magnitude of the surface current speed $U_0$, as noted in
section \ref{model}\ref{model1}\ref{model2}.
Figure \ref{f8}
shows a comparison of the values of $U_0/u_*$ calculated from the model 
(corresponding to
the horizontal portions of the curves in Fig. \ref{f7}) with the values 
that can be either obtained directly from Table 1 of \cite{Cheung_Street_1988},
or obtained from the data point with the largest depth in the datasets for 
each wind speed in Fig. \ref{f7}. It can be seen that the agreement is 
quite reasonable, with correlation coefficients of $\approx 0.95$ in both 
cases, although the model does
tend to systematically underestimate the data. However, given the strong 
assumptions adopted, 
the agreement is surprisingly good.

\section{Concluding remarks}
\label{conclusions}

This study presents a simple model for the wind-driven current existing in the
oceanic boundary layer in the presence of surface waves generated by the wind. 
The model sheds light
on two puzzling aspects that have been noted repeatedly about these currents,
for which a logarithmic profile model, with the friction velocity $u_*$ and 
roughness length $z_0$ as basic parameters, has often been adopted. Firstly,
if the current speed is scaled using the total friction velocity, measured 
independently, e.g., from the surface wind stress, the friction velocity 
diagnosed from shear in the current profile is smaller than expected,
being only a fraction of the total friction velocity. 
Secondly, the roughness length diagnosed from the same 
fitting procedure is much larger than expected, by various order of magnitude,
being inconsistent with the roughness length that would be estimated either
for an aerodynamically smooth flow, or aerodynamically rough flow affected 
by waves. The corresponding Charnock parameter appears to be enormously
amplified \citep{Bourassa_2000}. 

Both of these features are explained here as resulting from
a partition of the total turbulent shear stress into a shear-induced
component and a wave-induced component, which result from the local
mechanical production of this stress by the mean shear in the current 
profile, and by the Lagrangian strain rate associated Stokes drift of the 
waves, respectively, when
the effect of non-breaking waves is included in the equations of motion via 
the Craik-Leibovich vortex force. In this framework, the 
wave-associated part of the shear stress is not a property of the wave itself,
as assumed by some authors, but is a stress created on the turbulence
that co-exists with the shear-induced stress, by Stokes drift straining
of turbulent vorticity into the streamwise direction (the assumed direction 
of both the mean current and the Stokes drift) \citep{Teixeira_2011a}. This 
is independent from any vertical mixing associated with
pre-existing turbulence, or turbulence injected into the water by
wave breaking.

It is likely that this mechanism associated with non-breaking waves acts in
concert with other mechanisms related to wave breaking, and with the
transport of turbulence by itself in general, but the fact that it can 
account for the two phenomena mentioned above, and that its dependence on the 
turbulent Langmuir number appears to be confirmed by measurements, support
its relevance.

The model predicts that the part of the turbulent shear stress induced by
shear in the surface layer becomes a progressively smaller fraction of the
total stress near the surface and down to a depth of the order the wavelength 
of the waves as $La_t$ decreases. This leads to the perceived reduction of
the friction velocity. The model also predicts that the roughness length 
inferred if the uppermost portion of the current profile is disregarded 
is amplified
by various orders of magnitude as $La_t$ decreases, and that it scales with
$k_w^{-1}$, i.e., with the wavelength of the waves, at small $La_t$. 
The profile of the 
wind-induced current becomes flatter (that is, less different from its surface 
value) as $La_t$ decreases, which corresponds to an increase in wind speed.  

If the parameters in the model are adjusted appropriately (presumably to
account for the facts that there is no substantial wave breaking and the 
waves are not monochromatic), good agreement is found with the laboratory
measurements of \cite{Cheung_Street_1988}, which appear to be the only dataset 
that is precise and comprehensive enough for this purpose. Other more
recent datasets (\citealp{Siddiqui_Loewen_2007}, \citealp{Longo_etal_2012})
either seem unreliable, or do not provide complete enough information about 
the characteristics of the wave field or of the total shear stress. 

As the present one, a recent study of \cite{Sinha_etal_2015} uses insights
from \cite{Teixeira_2012} to develop a turbulence closure that includes wave
effects. However, the dataset they use to test their model, from LES of 
\cite{Tejada-Martinez_etal_2013}, refers to shallow water flow, and is
thus strongly
affected by the bottom boundary layer. \cite{Sinha_etal_2015} primarily 
focus on an analysis
of the current profile in wall-coordinates within the bottom boundary layer, 
but the full-depth current profiles shown by them (e.g., their Figs. 19 and 21) 
suggest a relatively modest agreement between their model in the top 
boundary layer adjacent to the air-water interface, despite the fact that they
include a term in the shear stress definition that is not local, accounting
for turbulent transport of TKE (which is not done here).

In order to bring the model presented here closer to real oceanic conditions,
and thus increase its usefulness, it is probably not only necessary to account 
for non-local mixing (which is important in some datasets), 
but also for the effect
of the Earth's rotation, as wind-driven currents are known to be typically
misaligned with the surface stress and rotate with depth, in accordance with 
Ekman layer theory. However, within the surface layer where the shear stress 
is the
primary mechanism shaping the current, shear at least is necessarily aligned 
with the wind stress, and thus the model presented here may still be directly
applicable to the streamwise component of the current.

\bibliographystyle{ametsoc2014}
\bibliography{references}

%

%


\end{document}